# Electrically tunable orbital coupling and quantum light emission from O-band quantum dot molecules

P.S. Avdienko,[1,2,*] L. Hanschke,[3,2] Q. Buchinger,[4] N. Akhlaq,[1,2] I. Lubianskii,[5] E. Weber,[5] H. Riedl,[5] M. Kamp,[4] T. Huber-Loyola,[4,6,7] S. Höfling,[4] A. Pfenning,[4,6] K. Müller,[3,2] and J.J. Finley[1,2,†]

[1]*Walter Schottky Institute, TUM School of Natural Sciences, Technical University of Munich, Am Coulombwall 4, 85748 Garching, Germany*
[2]*Munich Center for Quantum Science and Technology (MCQST), 80799 Munich, Germany*
[3]*Center for Quantum Engineering (ZQE) and School of Computation, Information and Technology, Technical University of Munich, 85748 Garching, Germany*
[4]*Julius-Maximilians-Universität Würzburg, Physikalisches Institut, Lehrstuhl für Technische Physik, Am Hubland, 97074 Würzburg, Germany*
[5]*Walter Schottky Institute, Technical University of Munich, Am Coulombwall 4, 85748 Garching, Germany*
[6]*Saqura Technologies GmbH, Paradiesstr. 15, 97855 Homburg am Main, Germany*
[7]*Karlsruhe Institute of Technology, Institute of Photonics and Quantum Electronics IQST, Engesserstr. 5, 76131 Karlsruhe, Germany*

We present the observation of electrically tunable quantum coupling of orbital states in individual InAs/InGaAs quantum dot molecules emitting in the telecom O-band ($\sim$ 1.3 μm). By tuning the static electric field along the growth axis of the QD-molecule, we observe pronounced anticrossings between excitonic transitions and determine the dependence of the interdot electron tunnel coupling on the interdot separation. As the electric field applied along the growth axis of the QD-molecules increases, positively charged exciton complexes sequentially emerge in the time-integrated emission spectra due to electron escape from the system while holes remain trapped. Moreover, for strong pumping, biexciton emission from the O-band molecules is identified. We demonstrate single-photon emission from the InAs/InGaAs QD-molecule emitting around 1.3 μm with a $g^{(2)}(0) = 0.017 \pm 0.002$ and explore the impact of tuning orbital coupling on the second-order correlation function.

## I. INTRODUCTION

The development of single-photon sources (SPSs) based on semiconductor quantum dot (QD) and QD-molecule (QDM) nanostructures emitting in the telecommunications bands is essential for integration with fiber optics and scalable silicon nanophotonic platforms [1–13]. Entangled multi-photon states are a key resource for measurement-based photonic quantum communication and computation [14–16]. In contrast to single QDs, it has been theoretically shown that QDMs hosting two optically addressable spins can deterministically generate multi-photon states with the required two-dimensional entanglement structures [16].

Most investigations of semiconductor-based SPSs focus on (In,Ga)As QDs grown on GaAs substrates emitting at wavelengths < 1 μm [17–27]. However, fiber-based quantum photonic systems require operation in the low-loss telecom O- and C-bands centered around 1.3 and 1.55 μm, respectively [1–13]. Promising platforms for telecom quantum light emission include InAs QDs grown on InP or GaAs substrates [4, 5, 7–13]. Photonic resonators are needed to enhance photon extraction efficiency and boost entanglement fidelities [1]. While InAs/InP QDs intrinsically emit the telecom bands, their integration with high-reflectivity distributed Bragg reflectors (DBRs) is limited by the low refractive-index contrast of InGaAlAs/InP mirror stacks [3, 8]. In contrast, O- and C-band InAs QDs grown on GaAs substrates enable site-selective fabrication and straightforward integration with high-contrast Al(Ga)As/GaAs DBRs and other photonic antennas [1, 7].

For conventional InAs/GaAs Stranski–Krastanov (SK) QDs, the low-temperature (LT) emission wavelength typically does not exceed 1.2 μm [28]. A further redshift of the emission into the O-band can be achieved by overgrowing the QDs with a thin In(Al,Ga)As(N) strain-reducing layer (SRL) [29–32]. This reduces the effective bandgap of the fundamental exciton transition and has been used by several groups to demonstrate efficient O-band SPSs on GaAs substrates [31, 33, 34].

Here, we report the synthesis and direct observation of quantum coupling in O-band QDMs formed by vertically stacking pairs of InAs QDs separated by a thin GaAs tunnel barrier. In this approach, the lower QD layer acts as a stressor that creates preferential nucleation sites for dot formation in the upper layer [35]. Vertical alignment of the two dots and the strength of the tunnel coupling is governed by the thickness [36] and Al-content of the $Al_xGa_{1-x}As$ layer separating them [21–25]. We surround the O-band QDMs by AlGaAs barriers (see Fig. 1) and embed them into the intrinsic region of a vertical p-i-n diode to facilitate wide tunability of the exciton energy and orbital coupling via the DC Stark effect [37].

For conventional In(Ga)As/GaAs QDMs emitting around 930 nm, the two dot layers are typically separated by a GaAs barrier with a thickness of up to 10 nm [35]. However, their structural and optical properties differ substantially from those of O-band InAs QDs capped with an InGaAs SRL [28–31, 34]. Consequently, we optimize the SRL thickness to ensure vertical correlation of the QD layers and investigate the range over which the tunnel-coupling strength can be tuned. Coupling of orbital states within the QDM is evidenced by pronounced anticrossings in the optical spectra, arising from electron tunneling between the QDs. A progressive transfer of spectral weight towards increasingly *positively* charged excitonic complexes is observed with increasing electric field, indicating hole accumulation in the upper QD of the molecule. These signatures are reproducibly observed across multiple QDMs and for samples with different interdot separations. Our results establish clear structure–property relationships for O-band InAs QDMs incorporating an InGaAs SRL. Finally, we demonstrate quantum light emission around 1.3 µm originating from coupled excitonic states of the QDM and single-photon emission at 1.3 µm with a $g^{(2)}(0) = 0.017 \pm 0.002$ under CW excitation.

* pavel.avdienko@tum.de
† jj.finley@tum.de

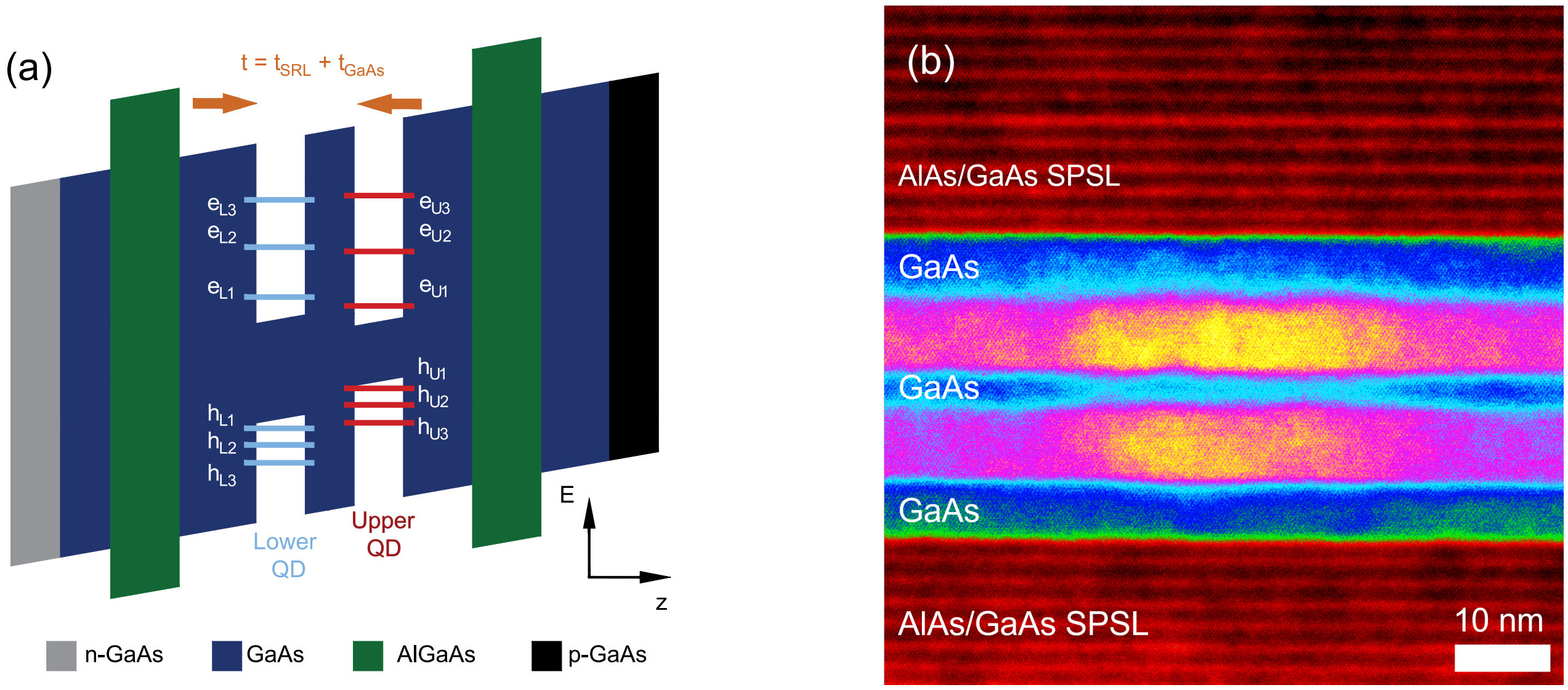

FIG. 1. (a) Schematic of p-i-n diode band diagram with InAs/InGaAs QDM including doped layers and AlGaAs barrier. (b) A false-color cross-sectional high-angle annular dark-field scanning transmission electron microscopy (HAADF-STEM) image of a single InAs/InGaAs QDM with a 3 nm GaAs barrier.

## II. EXPERIMENT

We investigated four samples grown by molecular beam epitaxy (MBE): (i) a reference sample that contains only a single layer of InAs/InGaAs O-band QDs, and samples (ii) (iii) and (iv) containing two layers of QDs with interdot separations of 3 nm, 5 nm and 10 nm GaAs, respectively (see Fig. 1. All samples were grown under identical conditions described in Appendix A.

As depicted schematically in Fig. 1(a), the InAs/InGaAs QDMs were embedded near the center of the p–i–n diode and surrounded by AlGaAs barriers to tune electric fields over $0-170$ kV/cm without carrier tunneling escape occurring [37]. The first InAs/InGaAs QD layer (lower) was deposited 5 nm above the 68-nm-thick barrier consisting of a short-period superlattice (SPSL) equivalent to $Al_{0.75}Ga_{0.25}As$ [31, 37], while the second QD layer (upper) was capped with a 5 nm GaAs spacer followed by an identical SPSL barrier. The thickness of the undoped GaAs layers located between the p- and n-type GaAs contact layers and the SPSL barriers was varied to maintain the same total diode lengths. The undoped GaAs layer thickness of sample (i) is 85 nm. Fig. 1(b) shows a representative high-angle annular dark-field scanning transmission electron microscopy (HAADF-STEM) image of QDM sample (ii), presented in a false-color filter to enhance contrast, revealing the SPSL barriers and the vertically stacked QDM.

Micro-photoluminescence (µPL) measurements at LT were performed on all samples as a function of applied voltage (PLV) subject to non-resonant continuous-wave (CW) excitation at 895 nm. The emission from the InGaAs SRL was observed around 1000 nm at 4 K. For the statistical analysis of tunnel coupling energy ($\Delta_{tc}$), we extended a scanning hyperspectral imaging technique (HSI) [2, 38] also to enable voltage-dependent HSI imaging. Each scan yields a four-dimensional dataset comprising PLV spectra at each spatial position (x, y), providing statistics across multiple individual QDMs.

## III. RESULTS AND DISCUSSION

Fig. 2(a) shows typical PLV spectra recorded from the reference sample (i) with a single QD layer. This data was acquired from a region of the substrate having a density of >1 QD/µm$^2$. As expected, all emission lines 1–4 exhibit a pronounced quantum-confined Stark effect (QCSE) as the electric field is increased, characterized by

$$\Delta E^{\mathrm{dir}}_{\mathrm{Stark}} = E^{\mathrm{dir}}(F) - E^{\mathrm{dir}}(F=0) = -p \cdot F - \beta \cdot F^2,$$

where $F$ is the static electric field strength, $p$ is the permanent exciton dipole moment, and $\beta$ is the exciton polarizability [37, 39]. The static electric field is $F = (V_{\mathrm{bi}} - V)/d$, where $V_{\mathrm{bi}} \approx 1.5$ V is the built-in voltage of the diode and $d = 334$ nm is the total thickness of the intrinsic region. The PLV data presented in Fig. 2(a) shows a >10 meV QCSE of two different InAs/InGaAs QDs with little quenching of the emission intensity due to the AlGaAs barriers. Parabolic fits to the different transitions yield a permanent excitonic dipole moments of $p/e = 0.52$ nm and polarizabilities of $\beta = 0.27\ \mu\mathrm{eV\,kV^{-2}\,cm^2}$ for lines 3 and 4, and $p/e = 0.35$ nm and $\beta = 0.16\ \mu\mathrm{eV\,kV^{-2}\,cm^2}$ for the most intense lines 1 and 2, respectively. The exciton dipole moment $p$ reflects the spatial separation between the centers of the electron and hole envelope wave functions in the absence of an external electric field. All four fitted bright emission lines, labeled 1–4, exhibit $p/e > 0$, indicating that the electron wave function is located below the hole wave function at $F = 0$ kV/cm, consistent with previous findings [40]. The polarizability is expected to be strongly dependent on the QD height [39] with $\beta \propto h_z^4$, giving rise to a substantially stronger parabolic Stark shift, by a factor up to 7, for InAs/InGaAs QDs compared to In(Ga)As/GaAs QDs.

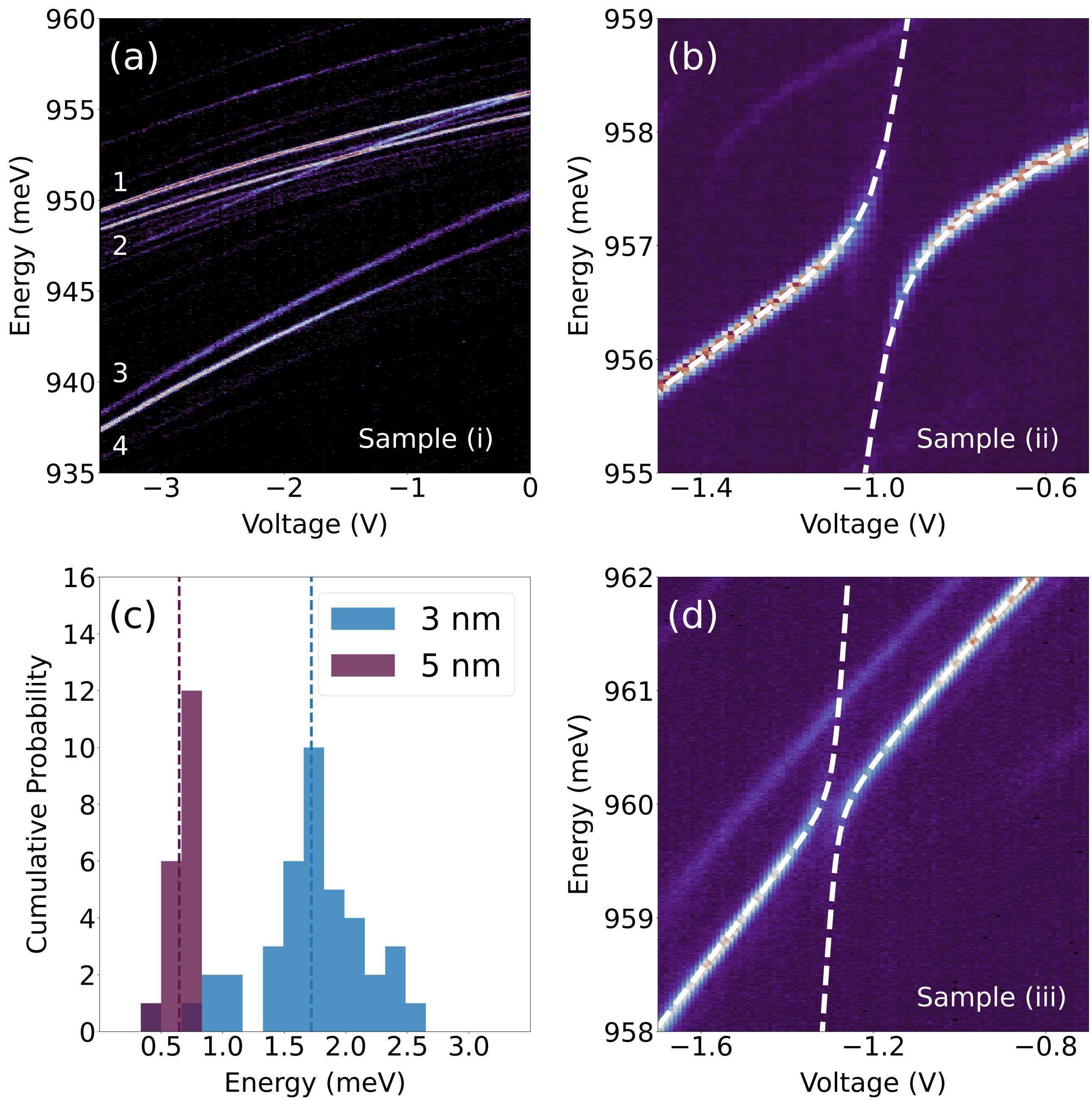

FIG. 2. (a), (b), and (d) PLV data recorder from a reference InAs/InGaAs QD sample (i) and InAs/InGaAs QDM samples (ii) and (iii) at T = 4 K under the CW 895 nm excitation with $P_{exc}$ = 1 μW. (b) and (d) The evolution of a representative anticrossing for GaAs barrier thickness $t_{\mathrm{GaAs}}$ of 3 nm and 5 nm, respectively. The anticrossing fit is shown as dashed lines. (c) Cumulative distributions of the tunnel-coupling energies $\Delta_{tc}$ for $t_{\mathrm{GaAs}} = 3$ nm and 5 nm. The mean values of $\Delta_{tc}$, 1.72 and 0.65 meV, are indicated by dashed lines.

Figs. 2(b) and 2(d) present two representative anticrossings (ACs) observed in the weak excitation regime for the InAs/InGaAs QDM samples (ii) and (iii). As the applied voltage is tuned, the ACs emerge due to coupling between spatially direct excitons ($X_{\mathrm{dir}}$), where the electron-hole pair is confined within the same QD, and indirect excitons ($X_{\mathrm{ind}}$), where the electron and hole reside in different QDs of the QDMs. $X_{\mathrm{dir}}$ follows the conventional QCSE, as described for sample (i), while the Stark shift of $X_{\mathrm{ind}}$ is given by $\Delta E^{\mathrm{ind}}_{\mathrm{Stark}} = e \cdot s_{\mathrm{ind}} \cdot F$, where $s_{\mathrm{ind}}$ is the center-to-center interdot separation. This strong linear Stark shift enables tuning of $X_{\mathrm{ind}}$ into resonance with $X_{\mathrm{dir}}$, resulting in an AC at the critical electric field $F_{\mathrm{AC}}$ due to electron tunnel coupling [35, 41]. At the resonance, the electron ($e$) or hole ($h$) components of the excitonic wave function hybridize to form molecular-like states [35, 41]:

$$E_{\pm} = E_1^0 + \frac{1}{2}\Delta \pm \frac{1}{2}\sqrt{\Delta^2 + (2t)^2}$$

with $\Delta = E_2^0 - E_1^0$ being the energy difference between the uncoupled states. Considering identical polarizabilities and intrinsic dipole moments, $\Delta = e \cdot p_{ind} \cdot (F - F_{AC})$, where $e \cdot p_{ind}$ represents the equivalent dipole moment of the $X_{\mathrm{ind}}$. The dashed lines on Fig. 2(b) and 2(d) show fits to the data using the energy splitting of the anticrossing, which is then

$$\Delta_{tc} = \sqrt{[e \cdot p_{ind} \cdot (F - F_{AC})]^2 + (2t)^2}.$$

At resonance, the two hybridized excitons are split by a tunnel-coupling energy $\Delta_{tc} = 2t$, giving rise to the experimentally observed AC. In this regime, the direct and indirect excitons have identical oscillator strengths.

Spin-spin exchange interactions in a charged QDM are primarily determined by $\Delta_{\mathrm{tc}}$ [22, 42], which can be tuned by varying the tunnel barrier thickness $t_{\mathrm{GaAs}}$ between the QD layers. To statistically investigate the dependence of $\Delta_{\mathrm{tc}}$ on the interdot barrier thickness $t_{\mathrm{GaAs}}$ in our QDMs, we performed μPL measurements on InAs/InGaAs QDM samples (ii), (iii) and (iv) with different $t_{\mathrm{GaAs}} = 3$, 5, and 10 nm, respectively. For this, we used the HSI technique [2] extended with a bias-voltage sweep at each scan step. As $t_{\mathrm{GaAs}}$ increases from 3 to 5 nm, the mean value of $\Delta_{\mathrm{tc}}$ decreases from 1.72 to 0.65 meV, indicted by dashed lines in Fig. 2(c). No anticrossings could be observed for sample (iv) with $t_{\mathrm{GaAs}} = 10$ nm, indicating that the vertical correlation between the pairs of QDs forming the molecules was lost. The behavior of an excitonic two-particle wave function in a strained, symmetry-broken QDM system is essentially nontrivial [43]. Nevertheless, within single- and many-particle descriptions of tunnel-coupled systems [43, 44], the electron tunneling rate $T \propto \Delta_{tc}$ at the given applied electric field can still be qualitatively described as being mostly dependent on the height and width of the potential barrier separating the QDs. Consequently, an increase in either the interdot barrier height (e.g., via a higher Al content in the barrier separating the QDs) or the barrier thickness, as well as deeper localization, leads to a suppression of tunneling, and consequently $\Delta_{\mathrm{tc}}$. HSI measurements acquired from $20 \times 20$ μm$^2$ areas with a QD density as low as 0.2 QD/μm$^2$ revealed 37 and 16 anticrossings for samples (ii) and (iii), respectively. Based on these data, the probability of observing anticrossings was estimated to be 74% and 32% for the QDM samples with $t_{\mathrm{GaAs}} = 3$ and 5 nm, respectively. The absence of ACs for a total lower wetting layer (WL) to upper WL separation of $t_{\mathrm{GaAs}} + t_{\mathrm{SRL}} = 17$ nm (see Fig. 1(a)) confirms that the tunnel coupling decreases sharply with increasing barrier thickness, occurring more rapidly in QDM systems with enhanced carrier confinement compared to In(Ga)As/GaAs emitting around 930 nm.

The panels (a) and (b) of Fig. 3 present PLV data recorded from sample (ii) in a broad range of applied bias when excitation is weak (300 nW) and strong (5 μW), respectively. For weak excitation, the formation of multi-exciton species is suppressed, and the majority of the emission lines stem from single exciton species populating the QD-molecule. At low electric fields ($< 70$ kV/cm), a dominant spectral line, marked $X$, exhibits a QCSE similar to that of a single QD. In contrast to the single QD, as the electric field is tuned into the range 70 to 85 kV/cm, we observe AC, labeled $AC1$, in the spectrum. Fitting yields $\Delta E_{\mathrm{dir/ind}}(F = 0) = |E_0^{X_{\mathrm{ind}}} - E_0^{X_{\mathrm{dir}}}| \approx 74$ meV, along with $s_{\mathrm{ind}} \approx 9.9$ nm, which is in agreement with the nominal WL–to–WL layer separation $t = t_{SRL} + t_{GaAs}$ of 10 nm. For electric fields $F > 85$ kV/cm, the electron–hole overlap for $X_{\mathrm{ind}}$ becomes negligible, and only the $X_{\mathrm{dir}}$ remains optically active in luminescence (see Fig. 3(a)).

As the electric field is further increased $\geq 98$ kV/cm, a series of abrupt steps are observed in the emission energy at specific voltages. They are denoted in Fig. 3(a) by the vertical dashed lines on the figure. At these steps, the dominant QDM emission lines quench and are sequentially replaced by other lines primarily shifted to higher energy. This characteristic behavior is indicative of sequential charging of the QDM and the dominant emission shifting to different charge states of the system. We note that optical excitation always generates equal numbers of photoexcited electrons and holes. As such, emission from an effectively positively charged complex in the molecule requires that electrons transfer out of the QD, being trapped against the lower AlGaAs barrier. We label the emission in these regions $X^{n+}$, where $n = 1 \rightarrow 5$ and tentatively identify the different features as arising from effectively *positively* charged excitons in the upper QD of the molecule. Hereby, the axial electric field results in preferential hole transfer into the upper QD and the faster comparative tunneling escape results in hole accumulation.

We now support our identification of these features. As shown by Regelman *et al.* [45], the relative energy of charged and neutral exciton complexes reflects the interplay between: (1) the total exciton energy expressed as $\Delta E_X = E_{e1} - E_{h1} - J_{e1h1}$, where $E_{e1/h1}$ are the ground-state energies of the electron and hole, respectively, and $J_{e1h1}$ is the electron–hole pair binding energy; (2) The *Coulomb shift* ($V_{ij}$, with $i, j = e, h$) accounting for the difference between Coulomb repulsion and attraction terms between the recombining exciton and the spectator carriers; (3) The always negative *exchange shift* ($\Delta E_{\mathrm{exch}}$) responsible for the appearance of multiple spectral lines within the same voltage range. Open shells in the final state give rise to spin multiplets whose energies depend on the relative spin orientation of the confined carriers (e.g., the range less than -2.3 V in Fig. 3), where the doublet structure $X^{2+}$ results from different triplet or singlet final states [46]) and (4) The *correlation shift* ($\Delta E_{\mathrm{corr}}$) arising from variations in carrier–carrier correlations within the interacting multi carrier system confined in the QD. According to Coulomb interaction considerations, the electron wave function is more weakly localized than that of the hole due to its smaller effective mass. Therefore, the relative strength of Coulomb interactions scale as $|V_{ee}| < |V_{eh}| < |V_{hh}|$. Assuming a harmonic confinement potential and accounting for the effective masses, $|V_{hh}|+|V_{ee}|>2|V_{eh}|$ [39, 47]. Based on these considerations, the negatively charged exciton ($X^-$) is expected to exhibit a redshift, whereas the positively charged exciton ($X^+$) shows a blueshift relative to the neutral exciton ($X$). Furthermore, the magnitude of the $X^-$ shift is predicted to be smaller than that of $X^+$ in InAs/GaAs QDs [47].

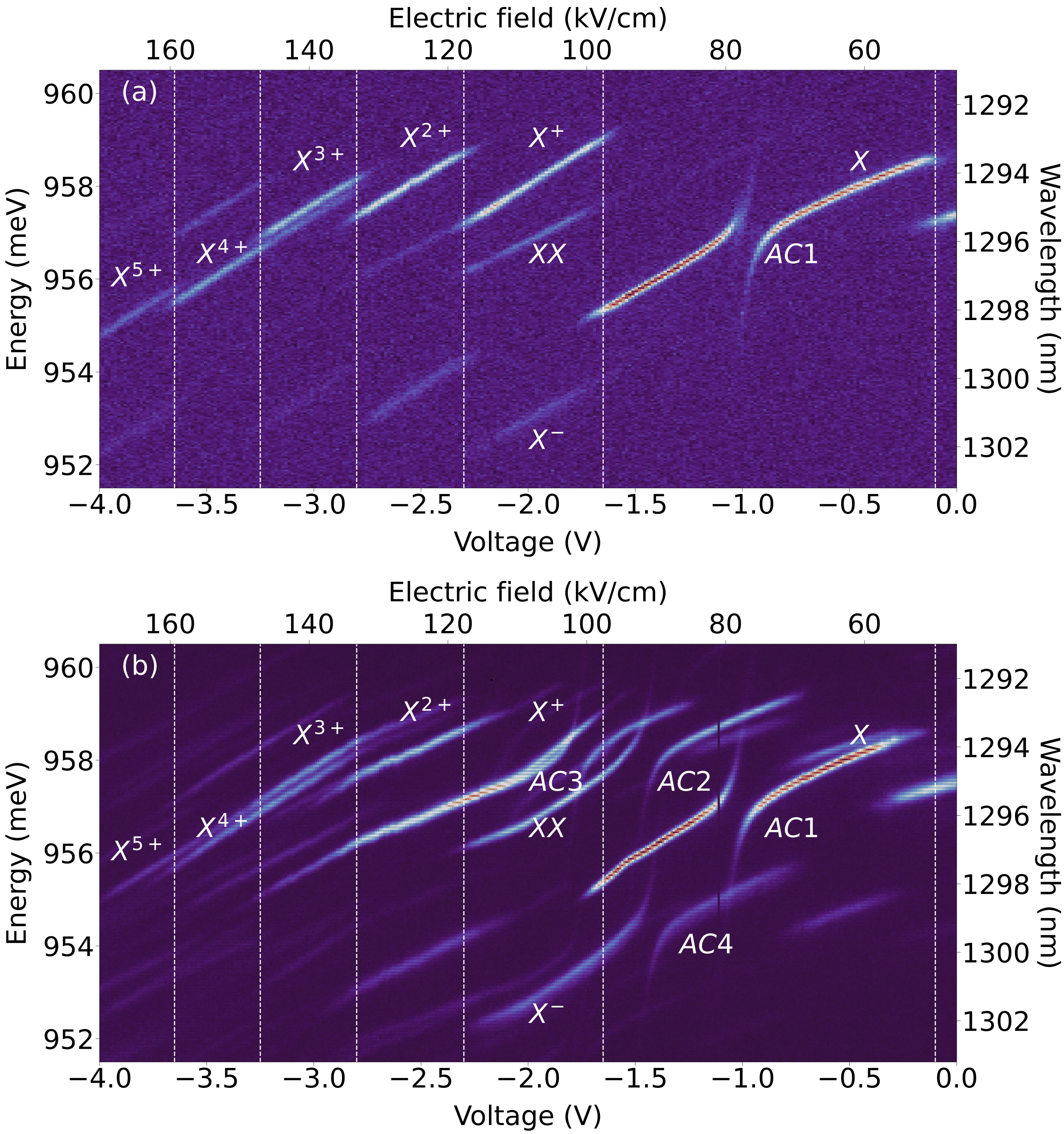

FIG. 3. PLV spectra of InAs/InGaAs QDM sample (ii) with $t_{GaAs}$ = 3 nm acquired at T = 4 K under CW 895 nm excitation with $P_{exc}$ = 300 nW (a) and 5 μW (b), respectively.

Taking the brightest line in each bias window, the experimentally observed energy shifts are:

$$X^+ - X \approx +3.65 \text{ meV},$$
$$X^{2+} - X^+ \approx +1.5 \text{ meV},$$
$$X^{3+} - X^{2+} \approx +0.85 \text{ meV},$$
$$X^{4+} - X^{3+} \approx -0.2 \text{ meV},$$
$$X^{5+} - X^{4+} \approx +0.2 \text{ meV}.$$

These values are qualitatively in agreement with the typical values reported for In(Ga)As/GaAs QDs [45, 46]. All our samples studied here were grown under identical growth conditions for the upper and lower QD layers, which almost unavoidably means that the upper QD is larger that confirmed by STEM (see Fig. 1(b)). Consequently, the upper QD has a stronger carrier confinement and polarizability in the coupled asymmetric QDM. Thus, we expect that at 0 V bias the electron/hole energy states satisfy $E_{e1}^{lower} > E_{e1}^{upper}$, while $E_{h1}^{lower} < E_{h1}^{upper}$ at $F < F_{AC1}^{X}$, as well as $\beta_{QD}^{upper} > \beta_{QD}^{lower}$ (see Fig. 1(a)). The observed charging behavior in Fig. 3(a) for the InAs/InGaAs QDM is in good agreement with the charging diagram of large, In-rich InAs/GaAs QDs emitting around 1.2 µm and embedded in an *n*–*i*–Schottky diode [47]. This correspondence further supports the picture that the lower InAs/InGaAs QD acts as a carrier reservoir for the upper QD. Moreover, the SRL overgrowth induces a redshift of the electron and hole states of InAs QD emission without significantly altering the hole-level splitting. At higher reverse bias, after reaching the neutral exciton anticrossing $F > F_{AC1}^{X}$ and $E_{e1}^{lower} < E_{e1}^{upper}$, while still $E_{h1}^{lower} < E_{e1}^{upper}$. Therefore, the positively charged complexes $X^{n+}$ blueshifted states are likely formed optically through the generation of an initial $X$ or biexciton ($XX$) in the upper QD, followed by electron (hole) tunneling to the lower (upper) QD.

Fig. 3(b) shows high-power PLV data. Increasing the excitation level leads to a pronounced modification of the PLV spectrum of the InAs/InGaAs QDM for electric fields $> 80$ kV/cm. Most strikingly, the emission lines exhibit abrupt discrete steps in the $X^+$-marked PLV region shown in Fig. 3(a) and evolve into pronounced features displaying additional anticrossings, labelled AC2-4 on the figure. Different charge states exhibit transitions with anticrossings that occur at different electric fields [48]. Anticrossings are not only limited to resonances between the lowest energy orbital states, but they can be used to optically probe the orbital states in each QD forming QDM. However, we note that the specific PLV anticrossing pattern of an excitonic state depends on couplings that occur in both initial and final states [48]. For example, the characteristic "X-shaped" anticrossing pattern that typically occurs for a trion ($X^\pm$) in a QDM [41, 48–50] is not observed if the initial and final states of the optical transition do not undergo resonance at similar electric fields. When the upper and lower QDs have significantly different confinement energies, characteristic X-patterns are not observed for $X^n$ states, depending on the preferential pathways of carrier relaxation and tunneling escape rates [49]. We note that the slope $p_{\text{ind}} = \Delta E/\Delta F$ has the same sign for AC1–AC4, indicating that the delocalization of the same component of the exciton wave function is responsible for all tunnel couplings.

We mapped the specific electric fields at which different ACs occur to the change of a few particle Coulomb interactions. Hereby, we assume that the differences between the resonance fields $F_{\text{AC}_\text{i}}$ (with $i = 1$–4) arise from Coulomb interactions between various orbital states, and express the shift of $F_{\text{AC}_\text{i}}$ relative to $F_{\text{AC}_1}$ to an energy offset using

$$\phi_{1\text{i}} = p_{\text{ind}} \left( F_{\text{AC}_\text{i}} - F_{\text{AC}_1} \right),$$

where $p_{\text{ind}} = e s_{\text{ind}} \approx e \cdot 9.9$ nm.

TABLE I. Summary of tunnel-coupling parameters

| AC | $\Delta_{tc}$ (meV) | $F_{\text{AC}}$ (kV/cm) | $\phi_{1\text{i}}$ (meV) |
|---|---|---|---|
| AC1 | 1.78 | 76.1 | 0 |
| AC2 | 2.1 | 91.6 | 15.3 |
| AC3 | 1.74 | 101.9 | 25.5 |
| AC4 | 1.8 | 90.4 | 14.0 |

The results of this analysis are summarized in Table I. We find that the tunnel coupling energies are similar for all anticrossings. As shown in Ref. [51], the *s*-*p* orbital splitting for electrons is expected to be considerably larger than for holes. The electron *s*-*p* splitting is typically several tens of meV, whereas for holes it is expected to be $\leq 10$ meV. The *p*-*d* splitting exhibits the opposite behavior: For holes, the magnitude of the splitting remains nearly constant at approximately 10 meV, while for electrons it changes nonmonotonically with height. The wave functions of the electron states are similar in shape to the conventional *s*-, *p*-, and *d*-like orbitals, whereas the hole states cannot be classified in this manner due to strong interband mixing [51]. Consequently, coupling between two electron states with different orbital characters, such as *s*-*p* or *s*-*d*, can occur through partial overlap of their in-plane wave functions belonging to different orbitally excited states [52]. However, such couplings are expected to exhibit reduced $\Delta_{tc}$ compared to, e.g., *s*-*s* or *p*-*p* coupling.

The power dependence of the integrated µPL intensity, $I \propto P_{\text{exc}}^{\alpha}$, measured in the range $-1.2\ V \rightarrow -1.3\ V$ reveals that the exponent $\alpha$ for AC1 is $0.95 \pm 0.01$ supporting our attribution of it as arising from a neutral exciton. In contrast, the exponents associated with AC2, AC3, and AC4 are $2.19 \pm 0.04$, $1.55 \pm 0.09$, and $1.59 \pm 0.08$, respectively. These values are indicative that the associated transitions are biexciton $XX$ and trion ($X^\pm$) states, respectively. A definitive interpretation of the nature of the multi-particle anticrossings cannot be made since shell filling effects become important as the excitation density increases towards the exciton saturation level $P_{\text{sat}}^{X}$, which occurs at a power 16.7 times higher than that in Fig. 3(a). Such shell filling effects are further enhanced by the presence of AlGaAs barriers (see Fig. 1(a)), which suppress carrier tunneling out of the QDM. Moreover, when the lowest energy orbital states energies become resonant at $E_{e1}^{\rm lower} < E_{e1}^{\rm upper}$ but $E_{e2}^{\rm lower} > E_{e2}^{\rm upper}$, the optical formation of $X^{n+}$ states from $X$ or $XX$ in the upper QD, followed by electron tunneling to the lower QD, becomes less straightforward, while hole transfer into the upper QD becomes more likely for $F > F_{\rm AC1}$. Therefore, AC2–AC4 are likely to reflect Coulomb-mediated hybridization of excitonic states in optically active InAs/InGaAs QDMs [53]. Such ACs have been shown to involve different excitonic states $|\psi_i\rangle$ having a unique configuration of single particle states and occur whenever a direct excitonic complex is tuned into resonance with an indirect one with $\Delta_{tc} \propto 2 \cdot t \cdot |\langle\psi_{dir}|\psi_{ind}\rangle|^2$, even in the absence of resonant single particle tunnelling. In the present case, after charge transfer between the two QDs, the stronger Coulomb binding energy of the resulting direct exciton state requires an additional electric field to restore resonance between the direct and indirect states, thereby forming the AC.

Based on the power dependence of the integrated $\mu$PL intensity, we tentatively identify AC2 as being a biexciton ($XX$) transition in the upper QD. The origin of AC4 may be partially explained by the involvement of a negatively charged trion ($X^-$) in the upper QD, with configuration $1e_s\,1e_p\,1h_p \rightarrow 1e_s$ [54], where $e/h$ denote electron/hole states in the $s$- or $p$-levels, respectively, accompanied by spin-conserving electron tunneling. The redshifted AC4 in Fig. 3(b) exhibits features absent in AC1–AC3. The $X^-$ trion consists of symmetric spin-triplet and antisymmetric spin-singlet configurations [54, 55]. The low-energy shoulder of the emission lines with predominantly direct character (in the 70–85 and 95–120 kV/cm ranges) vanishes near $F_{\rm AC4} = 90.4$ kV/cm. Furthermore, the splitting decreases monotonically from 120(20) $\mu$eV, disappearing as both the upper and lower branches acquire predominantly dark character. Similarly, the positively charged trion ($X^+$) contributes to the formation of AC3. The observed differences $\phi_{12} - \phi_{14} \approx 1.3$ meV and $\phi_{12} - \phi_{13} \approx 10.2$ meV are consistent with the variations in relative Coulomb interaction energies among these excitonic configurations.

Precise identification of charge states using photoluminescence excitation (PLE) spectroscopy is not presented here. However, to confirm that AC2-4 are indeed related to the charged excitonic states, we studied the power dependence of the second-order $g^2(\tau)$ intensity autocorrelation of direct excitonic states for voltages around AC1 and AC2 under non-resonant CW excitation at 895 nm. For these measurements, the voltage was fixed at -1.2V to avoid a change in tunneling rate. Distinguishing between different excitonic charge states using autocorrelation measurements can be carried out based on the character of correlation spectra, which is unique to each complex [56]. Fig. 4(a) exhibits the expected single photon behavior of a two-level system (InAs/InGaAs QDM) of the X state emitting at 956.7 meV under CW excitation, described, in the general case, by

$$g^{(2)}(t) = (1 - Ae^{-t/\tau_a}) \cdot (1 + Be^{-t/\tau_b}),$$

where $B$ and $\tau_b$ are set to zero in the absence of bunching. The fitting parameters are summarized in Table II in the Appendix.

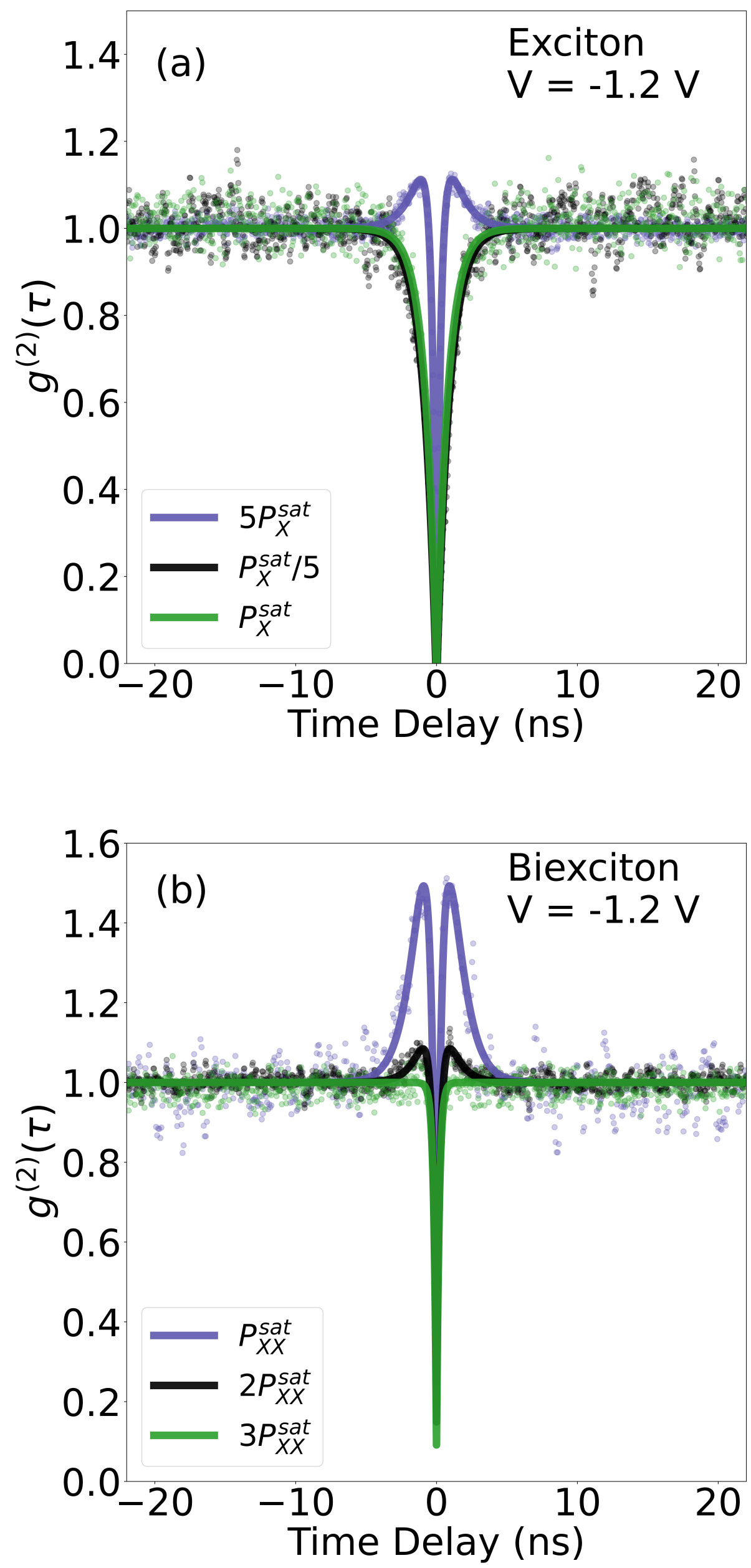

FIG. 4. Power-dependent second-order autocorrelation measurements and corresponding fits for the exciton (a) and biexciton (b) lines of sample (ii), recorded under 895 nm continuous-wave (CW) excitation at a bias of $-1.2$ V.

As long as the $X$ state does not reach intensity saturation (green and black experimental data in Fig. 4(a)), the correlation spectra exhibit a pronounced antibunching dip down to $g^{(2)}(0) = 0.017(2)$, with a width determined by the exciton lifetime. With increasing excitation power density, additional bunching at short time delays appears (lavender data and fit curve), decaying toward the average coincidence level over a timescale longer than that expected from pure radiative recombination. The magnitude of the bunching and the asso-ciated decay times are attributed to fluctuations in the charge configuration of the QD or QDM, caused by carrier shelving or Auger-type effects. Moreover, the $g^2(\tau)$ power dependence shown in Fig. 4(a) confirms that AC1 is the anticrossing of the neutral $X$.

The biexciton autocorrelation function of the line associated with the AC2 exhibits a markedly different behavior compared to the power dependence of the $X$ state. In this case, a strong characteristic bunching is observed at short time delays under weak CW excitation for the line at 958.5 meV, as presented in Fig. 4(b). The width of the dip confirms the fast decay rate of $XX$. Immediately after the detection of a biexciton photon, the upper QD is left with one e–h pair, which may either decay radiatively to the ground state or capture an additional carrier or an e–h pair to form $X^{\pm}$ or $XX$. As the excitation power decreases, the carrier capture rate is reduced, making it comparatively more probable to form a new $X^{\pm}$ or $XX$ at short timescales when the QD already contains one e–h pair. At longer delays, when the upper QD is typically empty, the formation of a new biexciton requires the capture of two e–h pairs. Consequently, the decay of the $XX$ bunching is expected to occur on a timescale comparable to the $X$ lifetime at $\leq P^X_{\mathrm{sat}}$, as can be seen in Fig. 4. The presence of tunnel coupling offers wide opportunities not only for controlling single-photon emission [23], but also for developing deterministic optical charging protocols in coupled QDMs within the telecom spectral range [16].

## IV. CONCLUSION

In summary, we have demonstrated the electrically tunable quantum coupling of orbital states in individual InAs/InGaAs quantum dot molecules emitting in around 1.3 µm. By tuning the applied electric field, we observed anticrossings of distinct excitonic transitions in a series of InAs/InGaAs QDM samples with $t_{GaAs}$ = 3 and 5 nm, using both HSI and standard PL techniques. The developed heterostructure design provides a platform for optical initialization, coherent manipulation, and readout of spin states in charged QDMs operating in the telecom O-band. Under strong excitation, biexciton emission from the O-band QDMs is identified. Finally, we have investigated the single-photon emission from InAs/InGaAs QDMs.

## ACKNOWLEDGMENTS

We gratefully acknowledge Beatrice Costa for assistance with the optical setup. Moreover, we gratefully acknowledge the Deutsche Forschungsgemeinschaft (DFG) via projects DIP (FI 947/6-1), MU4215/4-1 (CNLG), INST 95/1220-1 (MQCL) and INST 95/1654-1 (PQET), Germany's Excellence Strategy (MCQST, EXC-2111, 390814868) as well as the Bavarian Ministry of Economic Affairs (StMWi) via projects 6GQT and Munich Quantum Valley via NeQuS. In addition, we gratefully acknowledge the German Federal Ministry of Research, Technology and Space (BMFTR) via the projects PhotonQ (FKZ 13N15760 and 13N15759), QR.N (FKZ 16KIS2197 and 16KIS2209), and 6G-life. T.H.-L. acknowledges funding from the BMFTR via Qecs (FKZ 13N16272). A.T.P. acknowledges funding from the BMFRT via Ferro35 (FKZ 13N17641).

## DATA AVAILABILITY STATEMENT

The data that support the findings of this study are available from the corresponding author upon reasonable request.

## Appendix A

The investigated samples were grown by MBE on undoped 2-inch GaAs(001) substrates using a Veeco Gen II MBE system. The growth rates were 2 Å/s for GaAs, 1 Å/s for AlAs, and $\sim$ 0.15 Å/s for InAs. The indium cell temperature was held constant during the growth of both the InAs QD layer and the $In_{0.25}Ga_{0.75}As$ SRL. A gradient deposition of 1.7 monolayer (ML) (nominal coverage at the wafer center) of InAs was performed in the sub-ML regime using a sequence of InAs growth and As exposure cycles with a growth time of $t_{\mathrm{g}} = 3$ s and an interruption time of $t_{\mathrm{b}} = 3$ s at $T_S = 520$°C. The substrate was continuously rotated at 12 RPM after the initial four sub-ML deposition cycles to establish a controlled InAs thickness gradient across the wafer. For all samples, the InAs QDs were grown on the nominally flat GaAs surface at $P_{As}$(BEP) = $7 \times 10^{-6}$ Torr without a pattern-defining layer [57]. Following the growth of the first QD layer, $T_S$ was reduced to 490 °C during a 90 s interruption under As flux before overgrowth with a nominally 7-nm-thick $In_{0.25}Ga_{0.75}As$ SRL where $P_{As}$(BEP) was increased to $1 \times 10^{-5}$ Torr. After capping the first InAs/InGaAs QD layer with 2 nm of GaAs, $T_S$ was linearly ramped back to 580 °C to deposit an additional 1–8 nm GaAs barrier layer, ensuring a smooth surface for subsequent growth of the second InAs/InGaAs QD layer.

Diode mesas with an area of $400 \times 400$ µm$^2$ were defined using maskless lithography and wet etched down to the n-doped GaAs layer. Standard electrical contacts were fabricated by depositing a Ge/Au/Ni/Au metal stack onto the n-GaAs, followed by annealing at 420 °C, and a Ti/Pt/Au metal stack onto the p-GaAs without an annealing step.

TABLE II. $g^{(2)}(t)$ fitting parameters

| **Power** | | $\tau_a$ (ns) | $\tau_b$ (ns) |
|---|---|---|---|
| $P_X^{sat}/5$ | 0.04(4) | 1.1(1) | 0 |
| $P_X^{sat}$ | 0.017(2) | 0.89(1) | 0 |
| $5P_X^{sat}$ | 0.010(2) | 0.28(1) | 2.2(1) |
| $P_{XX}^{sat}$ | 0.18(8) | 0.50(13) | 1.23(17) |
| $2P_{XX}^{sat}$ | 0.13(2) | 0.28(2) | 1.09(14) |
| $3P_{XX}^{sat}$ | 0.09(2) | 0.18(1) | 0 |